\documentclass[aps,prb,superscriptaddress,twocolumn,oneside,floatfix,amsmath,showpacs,amssymb]{revtex4}
\usepackage{graphicx}

\begin{document}

\title{Common Fermi-liquid origin of T$^2$ resistivity and superconductivity in n-type SrTiO$_3$}

\author{D.~van~der~Marel}
\author{J.~L.~M.~van~Mechelen}
\affiliation{D\'{e}partement de Physique de la Mati\`{e}re Condens\'{e}e, Universit\'{e} de Gen\`{e}ve, CH-1211 Gen\`{e}ve 4, Switzerland}
\author{I.~I.~Mazin}
\affiliation{Center for Computational Materials Science, Naval Research Laboratory, Washington D.C., USA.}

\date{\today}

\begin{abstract}
A detailed analysis is given of the T$^2$ term in the resistivity observed in electron-doped SrTiO$_3$. Novel bandstructure data are presented, which provide values for the bare mass, density of states, and plasma frequency of the quasiparticles as a function of doping. It is shown that these values are renormalized by approximately a factor 2 due to electron-phonon interaction. It is argued that the quasiparticles are in the anti-adiabatic limit with respect to electron-phonon interaction. The condition of anti-adiabatic coupling renders the interaction mediated through phonons effectively non-retarded. We apply Fermi-liquid theory developed in the 70's for the $T^2$ term in the resistivity of common metals, and combine this with expressions for $T_c$ and with the Brinkman-Platzman-Rice (BPR) sum-rule to obtain Landau parameters of n-type SrTiO$_3$. These parameters are comparable to those of liquid $^3$He, indicating interesting parallels between these Fermi-liquids despite the differences between the composite fermions from which they are formed.
\end{abstract}

 \maketitle

\section{Introduction\label{sec:introduction}}
SrTiO$_3$ is a semiconductor which, when doped with a low density of electrons, becomes a good conductor with relatively high mobility and strong temperature dependence of the electrical resistivity and the infrared optical conductivity. At low temperatures the material becomes superconducting\cite{schooley1967} with a maximum reported T$_c$ of 1.2 K\cite{bednorz1988}, although superconductivity is usually reported below 0.7 K with a dome-shaped doping dependence of T$_c$\cite{koonce1967,binnig1980}. Superconductivity is also observed below 0.3 K in the two-dimensional electron-gas formed at the interface between SrTiO$_3$ and LaAlO$_3$\cite{reyren2007} where the carrier-concentration dependence of T$_c$ has also a dome shape\cite{caviglia2008}. The DC resistivity below 100 K has a T$^2$ temperature dependence, which has been attributed to electron-electron scattering by some groups\cite{tokura1993,okuda2001,hussey2005}. On the other hand, resistivity of the form $\rho(T)\propto 1/\sinh^2{(\omega_0/2T)}$, which is almost $T^2$-like, was found in La$_{1-x}$Ca$_x$MnO$_3$ \cite{zhao2000} and in doped LaTiO$_3$ \cite{gariglio2001} in accordance with the expected behavior of small polarons\cite{bogolomov1968}. However, n-type SrTiO$_3$ appears to be described well by the model of large polarons with a Froehlig-type electron-phonon interaction\cite{devreese2010}. Conditions in this material are therefore rather remote from those addressed by the small-polaron model\cite{bogolomov1968}, and the question as to why the T$^2$ behaviour dominates up to high temperature remains as yet open.

The resistivity near absolute zero has been known to be of the form $\rho=AT^2$ in platinum \cite{dehaas1934} and other transition metal elements\cite{white1967,schwerer1968,anderson1968,white1959,schriempf1968}, with $A$ ranging from $2.5\cdot 10^{-6} \mu\Omega$cmK$^{-2}$ (osmium) to $A=3 \cdot 10^{-5} \mu\Omega$cmK$^{-2}$ (palladium). M. J. Rice has explained these observations in terms of the Baber mechanism\cite{baber1937,rice1968}.
T$^2$ resistivity was subsequently observed in the alkali metals (see Ref.~\onlinecite{bass1990} for a review), with $A=3\cdot10^{-6}\mu\Omega$cmK$^{-2}$ for Li\cite{krill1971,sinvani1981}, and an order of magnitude smaller values for K and Na\cite{vanKempen1976,levy1979,bass1990}. Based on the assumption that the Coulomb repulsion is the only interaction between electrons, Lawrence and Wilkins\cite{lawrence1973} calculated values in the range from $10^{-8}$ to $10^{-10}$ $\mu\Omega$cmK$^{-2}$ for the alkali-metals. MacDonald obtained similar values, and showed that the dominant contribution to the $T^2$ term in the resistivity results from phonon-mediated interactions \cite{macdonald1980,macdonald1981}.
A value several orders of magnitude higher, $A=0.02 \mu\Omega$cmK$^{-2}$, was observed for stoichiometric TiS$_2$\cite{thompson1975}, and the resistivity of Ti$_{1+x}$S$_2$ as a function of carrier concentration was observed to follow the relation $n^{-5/3}T^2$ in agreement with the theoretical expressions in Ref.~\onlinecite{lawrence1973}.

In 1968 M. J. Rice pointed out\cite{rice1968}, that the coefficient $A$ should vary predominantly as the square of the linear electronic specific heat coefficient $\gamma$; in particular he showed that the experimental data of elemental $3d$, $4d$ and $5d$ transition metals satisfy the relation $A/\gamma^2=4 \cdot 10^{-7} \mu\Omega$ cm (mole K /mJ)$^2$. Heavy fermion compounds are characterized by very large values of $A$ and $\gamma$. Kadowaki and Woods\cite{kadowaki1986} summarized the situation by showing that $A/\gamma^2$ in this group of materials is a factor $\sim 25$ larger than in aforementioned data of elemental transition metals. According to the theory of electron-electron scattering\cite{baber1937,rice1968,nozieres1999,lawrence1973} the ratio $A/\gamma^2$ contains indeed several non-universal factors, including the square of the strength of the effective electron-electron interaction. Since in general the interactions differ in nature from one group of materials to another, the same values of $A/\gamma^2$ are only expected within a particular group. The carrier density constitutes another non-universal factor, which is particularly significant for doped semi-conductors in view of their tunable carrier density. Hussey\cite{hussey2005} proposed therefore a re-scaling of the Kadowaki-Woods plot to account for, among other factors, variations in carrier density, and demonstrated that this notion is supported by the strong doping dependence of $A$ in hole-doped LaTiO$_3$.

Here we return to the possibility that the T$^2$ resistivity in n-type SrTiO$_3$ could be a consequence of a quasi-non-retarded interaction between dressed quasiparticles. The $A$-coefficents of SrTi$_{1-x}$Nb$_x$O$_3$, a few examples of which are listed in table \ref{table:fit}, are large. Since, as has been demonstrated by Thompson\cite{thompson1975}, $A\propto n^{-5/3}$, this is a natural consequence of the low carrier density. For example, SrTi$_{0.98}$Nb$_{0.02}$O$_3$ has a carrier density $n=3.4\cdot 10^{20}$ cm$^{-1}$, while lithium $n=4.7\cdot 10^{22}$ cm$^{-1}$. If we assume that everything else is the same for these two materials, the $A$ coefficient of SrTi$_{0.98}$Nb$_{0.02}$O$_3$ should be 4000 times larger than the one of Li. In reality they differ by a factor 8000, hence from this perspective the strength of the quasiparticle-quasiparticle scattering in SrTi$_{1-x}$Nb$_x$O$_3$ is not drastically different from that in lithium.

An obvious source of interaction in doped SrTiO$_3$ is provided by the overlap of the screening clouds surrounding the electrons provided by the interaction with the lattice.
The main phonons involved in this screening are optical ones, with the important consequence that their energy exceeds the Fermi energy for the doping levels where superconductivity is observed. The polaron-polaron interaction mediated by these phonons is then effectively non-retarded, an unconventional aspect which we consider to be crucial for the observed T$^2$ dependence of the relaxation rate.
The effective electron-electron interactions can also lead to the formation of Cooper pairs. Based on our analysis of the T$^2$ relaxation rate and of the superconducting transition temperatures we obtain an interaction of weak to moderate strength, making implausible scenarios where a substantial fraction of the charge carriers is paired in the normal state.
\begin{figure}[ht]
\begin{center}
\includegraphics[width=\columnwidth]{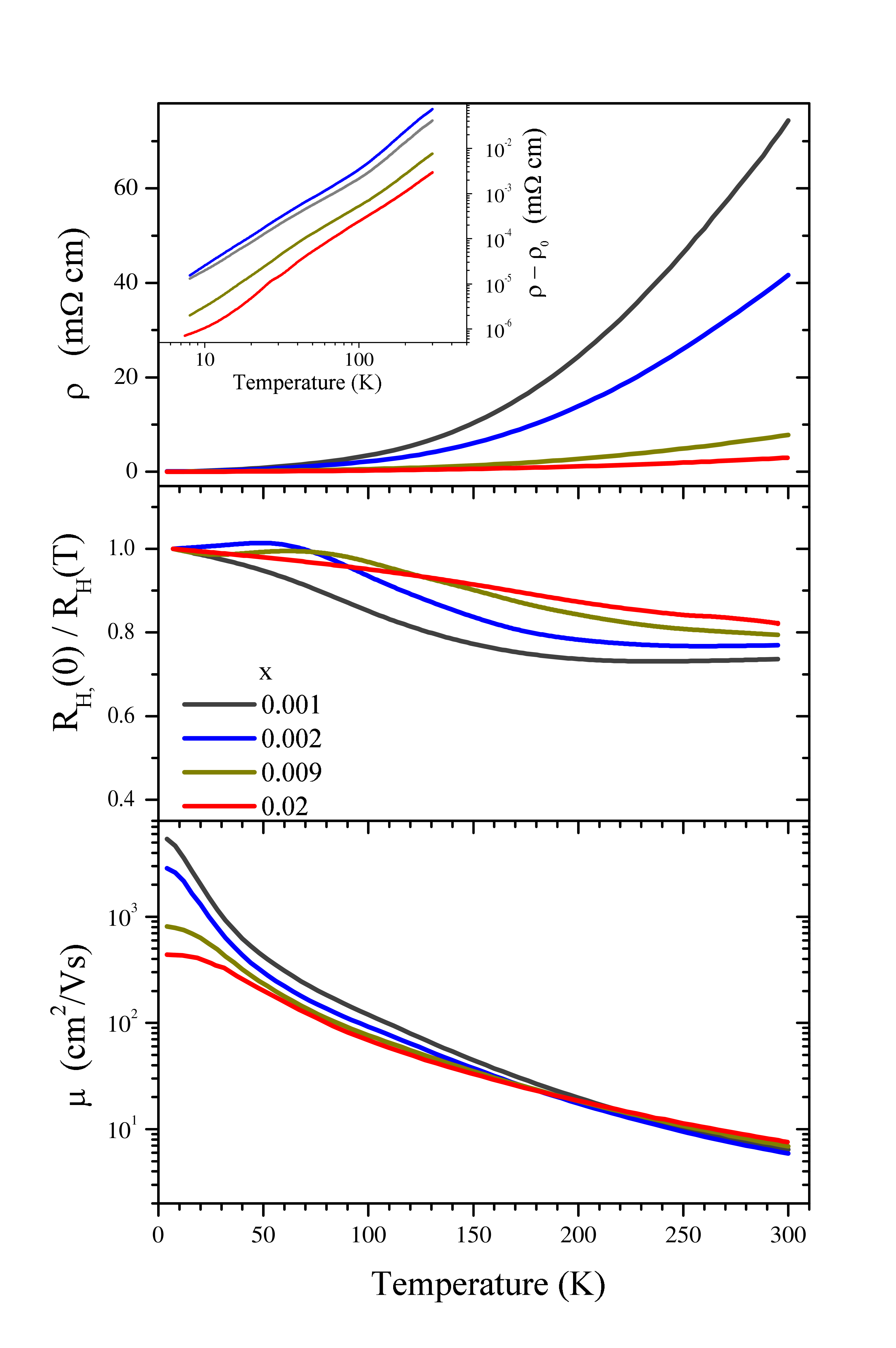}
\caption{Temperature dependence of the resistivity, the inverse Hall constant and the mobility of SrTi$_{1-x}$Nb$_x$O$_3$ for different carrier concentrations.
}
\label{fig:transport}
\end{center}
\end{figure}
\section{Transport properties\label{sec:transport}}
In Fig.~\ref{fig:transport} the transport data of SrTi$_{1-x}$Nb$_x$O$_3$ with different carrier concentrations are shown as a function of temperature\cite{thesisdook2010}. Hall data are presented as $R_{H,0}/R_H(T)$, where $R_{H,0}$ represents the zero temperature limit, for which the Hall charge carrier densities per unit cell $x_H=-a^3/(e R_{H,0})$ are $0.105$\%, $0.196$\%, $0.870$\% and $2.00$\%, which is within 12 $\%$ of aforementioned Nb concentrations specified by the supplier.
At 4 K we observe fairly high mobilities in the range from 400 to 6000 cm$^2$/Vs, which drop gradually as a function of increasing temperature to approximately 6 cm$^2$/Vs at room temperature. These high mobilities at cryogenic temperatures are the first indication that n-type SrTiO$_3$ is a clean Fermi-liquid of mobile charge carriers. Concentrating now on the temperature dependent properties, we take a closer look at the inverse Hall constants. First of all we notice that the sensitivity to temperature changes diminishes for increasing carrier concentrations. We consider the possibility that the system has multiple electron-type bands. The effective Hall density $n_H=-e/R_{H}$ of a multiband-band system with carrier density $n$ and fractional occupation of the j'th band $x_j$  with mobily $\mu_j$ is given by the expression $n_H/n=(\sum_j x_j\mu_j)^2/(\sum_j x_j\mu_j^2)\le 1$. The limiting case $n_H/n=1$ occurs when only one band is occupied, or/and if $\mu_j$ is independent of $j$. In all other cases $n_H/n<1$. The temperature dependence is well illustrated by the case where at $T=0$ only one band is occupied ($x_1=1$). Increasing temperature makes $x_j>0$ for $j\ge 2$ and $x_1<1$, consequently $n_H/n$ is reduced. When two or more bands are already occupied at $T=0$, the relative change in occupation number as a function of temperature is weaker and consequently $n_H$ will be less temperature dependent. In Ref.~\onlinecite{vanmechelen2008} a weak temperature dependent decrease of the Drude spectral weight, $\omega_p^2$, was reported for temperatures higher than 100 K. A gradual temperature induced transfer of part of the electrons to states with a higher effective mass (and consequently lower mobility) then provides a natural explanation for both phenomena: The temperature induces a decrease of $\omega_p^2$ because it is inversely proportional to the mass, and an increase of $R_{H}$.

\begin{figure}[ht]
\begin{center}
\includegraphics[width=\columnwidth]{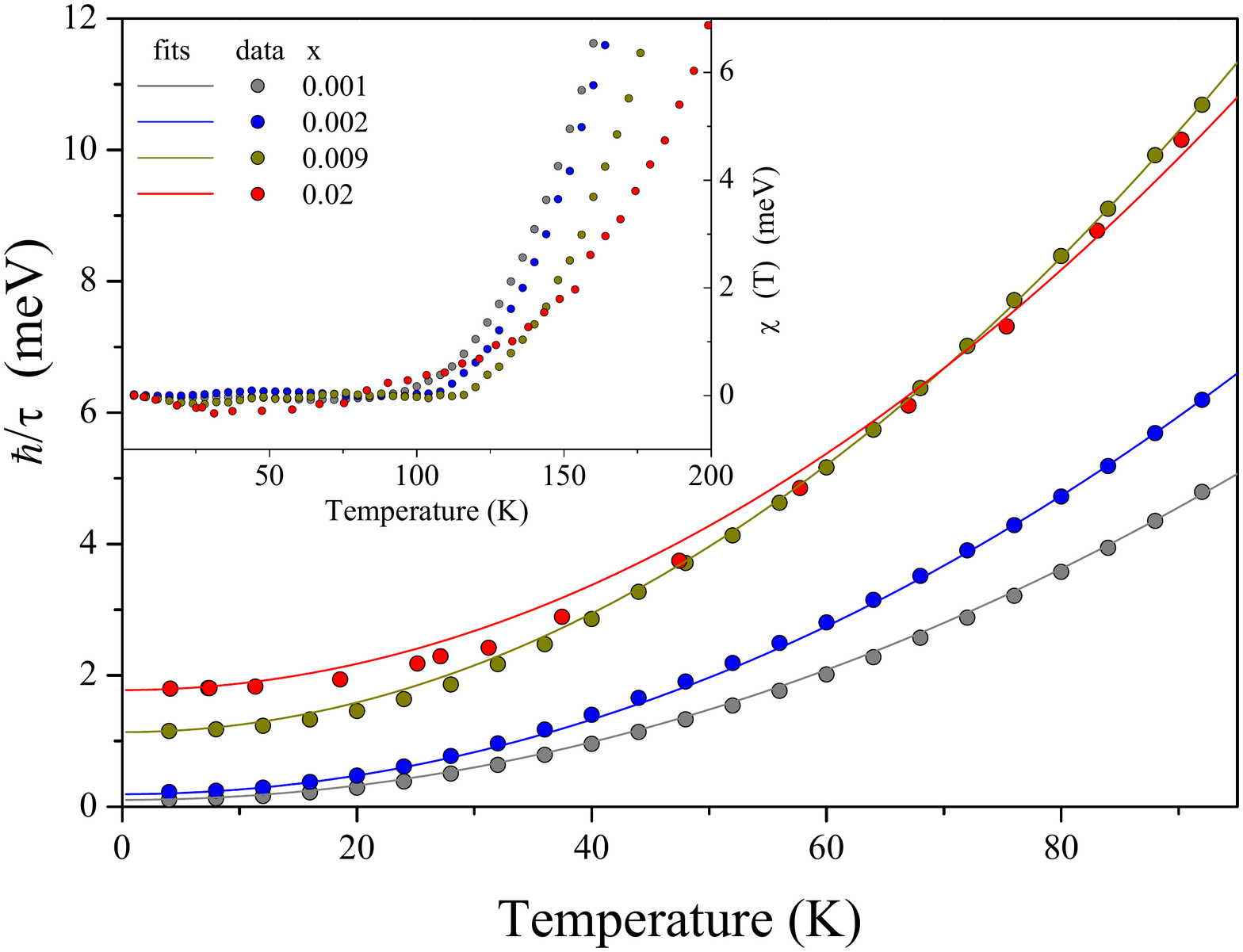}
\caption{Temperature dependent relaxation rates of SrTi$_{1-x}$Nb$_x$O$_3$ for 4 different carrier concentrations, using the relation $\rho(T)=4\pi\omega_p^{-2}\tau^{-1}$. Plasma-frequencies, $\omega_p$, are obtained from the Drude spectral weight measured with infrared spectroscopy\cite{vanmechelen2008} and listed in table~\ref{table:fit}. Inset: Difference between experimental data and fitted curve, $\chi(T)=\hbar/\tau(T)-\hbar/\tau_{fit}(T)$, demonstrating upward departure from $T^2$ behavior of the resistivity above 100 K.}
\label{fig:tau}
\end{center}
\end{figure}

The resistivities have a small residual component. The values for $\rho_0$ were determined by fitting the data below 15 K to a constant plus a power law, and these values of $\rho_0$ are used in the remainder of the analysis. The inset of Fig.~\ref{fig:transport} shows the resistivity, from which the residual component has been subtracted, on a double log scale, indicating a power law like increase as a function of temperature. For further analysis it is useful to convert the resistivities to relaxation rates using the expression
\begin{equation}
\rho(T) = \frac{4\pi}{\omega_p^2\tau}
\label{eq:resistivity}
\end{equation}
For $\omega_p^2$ we substitute the values measured with time-domain infrared spectroscopy on the same set of samples\cite{vanmechelen2008}.
The residual relaxation rate turns out to be proportional to $x$. Since $x$ is just the density of Nb$^{4+}$ ions, and these ions act as scattering centers, this (near) proportonality of scatttering to Nb-concentration is reasonable.
In the following discussion we will focus on the behavior of $\hbar/\tau$ below 100 K, where both $\omega_p^2$ and $R_{H}$ are independent of temperature.
The results of least-square fitting the relation $\hbar/\tau-\hbar/\tau_0=\alpha_{\eta}(T/100K)^{\eta}$, summarized in table~\ref{table:drude}, clearly demonstrate that the temperature dependence of the resistivity up to 100 K follows closely a $T^2$ power law. With this in mind we fitted $a_2$ in the expression $\hbar/\tau-\hbar/\tau_0=a_2T^2$, which values are listed in table~\ref{table:fit} and the corresponding fits are displayed together with $\hbar/\tau$ in Fig.~\ref{fig:tau}. Attempts to improve the fit in the 4-100 K range by adding a $T^3$ term decreases $\chi^2$, and affects $a_2$ somewhat. However, the prefactor of the $T^3$ term is negative for x=0.009, hence a $T^3$ term below 100 K gives unphysical results and should be dropped, with the only possible exception the x=0.02 sample. The difference between the data and the fit is constant upto 100 K, and grows rapidly at higher temperature, indicating that an additional component to the resistivity becomes active at that temperature. Such behavior is consistent with aforementioned interpretation of Hall data and spectral weight data, namely if electrons are transferred to lower mobility states, the resistivity will deflect upward from the trend observed at lower temperatures. We will return to this issue in the discussion of the mean free path in section~\ref{sec:manybody}.
\begin{table}[ht]
\begin{tabular}{cccccccc}
  \hline
$x_n$&$x_H$ & $\rho_0$& $\hbar\omega_p$& $\hbar/\tau_0$& $\alpha_{\eta}$ & $\eta$& $\chi^2$\\
	&&$\mu\Omega$cm&	meV&	meV       & meV     &	    &	meV$^2$\\
  \hline
    0.001&0.0011&62.4& 111& 0.104& 5.56& 2.09&0.00044\\
    0.002&0.0020&57.6& 157& 0.191& 7.04& 1.94&0.00044\\
    0.010&0.0087&53.0& 399& 1.135& 11.4& 2.04&0.0039\\
    0.020&0.020&41.8& 562& 1.776& 10.5& 2.25&0.0039\\
  \hline
\end{tabular}
\caption{First column: Nominal doping. Second column: Hall number in the zero temperature limit. Third column: Residual resistivity. Fourth column: Drude plasma frequency. Column 5: Residual relaxation rate. Columns 6 to 8: Fitting parameters of the temperature dependent relaxation rate fitted to a power law and corresponding variance.}
\label{table:drude}
\end{table}
\begin{table}[ht]
\begin{tabular}{ccccccc}
  \hline
$x$   &	a$_{2}$           &$\chi^2$  &	a$^{\prime}_2$ &	    a$_3$&$\chi^2$ & A                        \\
	  &$\mu$eVK$^{-2}$  & meV$^2$&$\mu$eVK$^{-2}$&neVK$^{-3}$&meV$^2$&$\mu\Omega$cmK$^{-2}$\\
  \hline
0.001 &	$0.55$            &0.0020    &	         $0.49$&	    $0.6$&	 0.0003& $0.33$\\
0.002	&	$0.71$&	0.0019&	$0.76$&	 $-0.6$&0.0006&$0.21$\\
0.009	&	$1.13$&0.0058&	$1.08$& $0.6$&	0.0047&$0.053$\\
0.02	&	$1.00$&0.0445&	$0.72$&	 $3.3$&	0.0076&$0.024$\\
  \hline
\end{tabular}
\caption{First column: Hall numbers rounded off to one significant digit, used in Figs.~\ref{fig:tau} and~\ref{fig:ioffe_regel} to label the samples. Second and third columns:  Fitting parameters and variance of the temperature dependent relaxation rate to a T$^2$ law. Fitting curves corresponding to $a_2$ are compared to the experimental data in Fig.~\ref{fig:tau}.
Columns 4 to 6: Fitting parameters and variance of the temperature dependent relaxation rate to a $T^2+T^3$ dependence.
Column 7: The $A$ coefficient in the relation $\rho=\rho_0+AT^2$.
The values of $\hbar/\tau_0$ are those of table~\ref{table:drude}.}
\label{table:fit}
\end{table}

\section{Band structure\label{sec:bands}}
SrTiO$_3$ has a cubic crystal structure at room temperature, which becomes tetragonal below a structural phase transition at 105 K. A 3 eV gap separates the filled oxygen $2p$ bands from the empty Ti $3d$ bands\cite{shanthi1998,cardona1965}. In Refs.~\onlinecite{vanmechelen2008,devreese2010,meevasana2010} we compared experiments to novel {\em ab initio} band calculations, the details of which have not been presented in the literature. In the present article we make again extensive use of the same new {\em ab initio} data. Since there are some differences compared to previously published bandstructure calculations, the new {\em ab initio} calculations are presented here in some detail.

First principles calculations were performed using the Linear Augmented Plane Wave method as implemented in the WIEN2k code~\cite{wien2k} and the generalized gradient approximation for the exchange-correlation potential in the form proposed by Perdew and coworkers~\cite{perdew} (See Appendix~\ref{appendix:A}.)
\begin{figure}[ht]
\begin{center}
\includegraphics[width=\columnwidth]{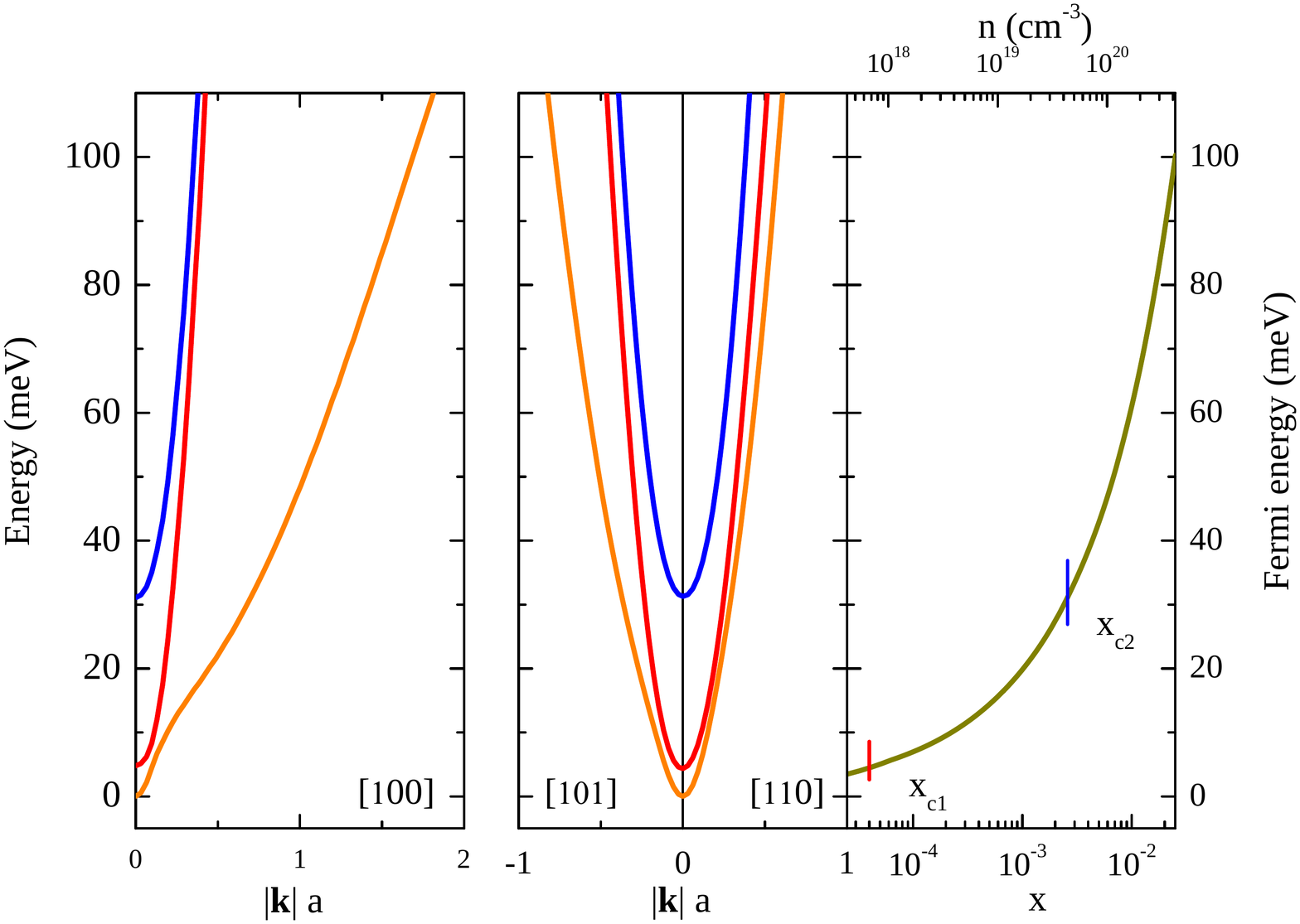}
\caption{Band dispersion of the lowest unoccupied bands of SrTiO$_3$ in the low temperature tetragonal phase. The directions in momentum space are labeled according to the high temperature cubic Brillouin zone, so that $[1,0,0]$ corresponds to momentum along the Ti-O bond direction. The rightmost panel indicates the position of the Fermi energy as a function of carrier concentration. $x_{c1}=4.0\cdot 10^{-5}$ and $x_{c2}=2.6\cdot 10^{-3}$ are critical carrier concentrations where the Fermi energy enters the second and the third band. }
\label{fig:dispersion}
\end{center}
\end{figure}
\begin{figure}[ht]
\begin{center}
\includegraphics[width=0.8\columnwidth]{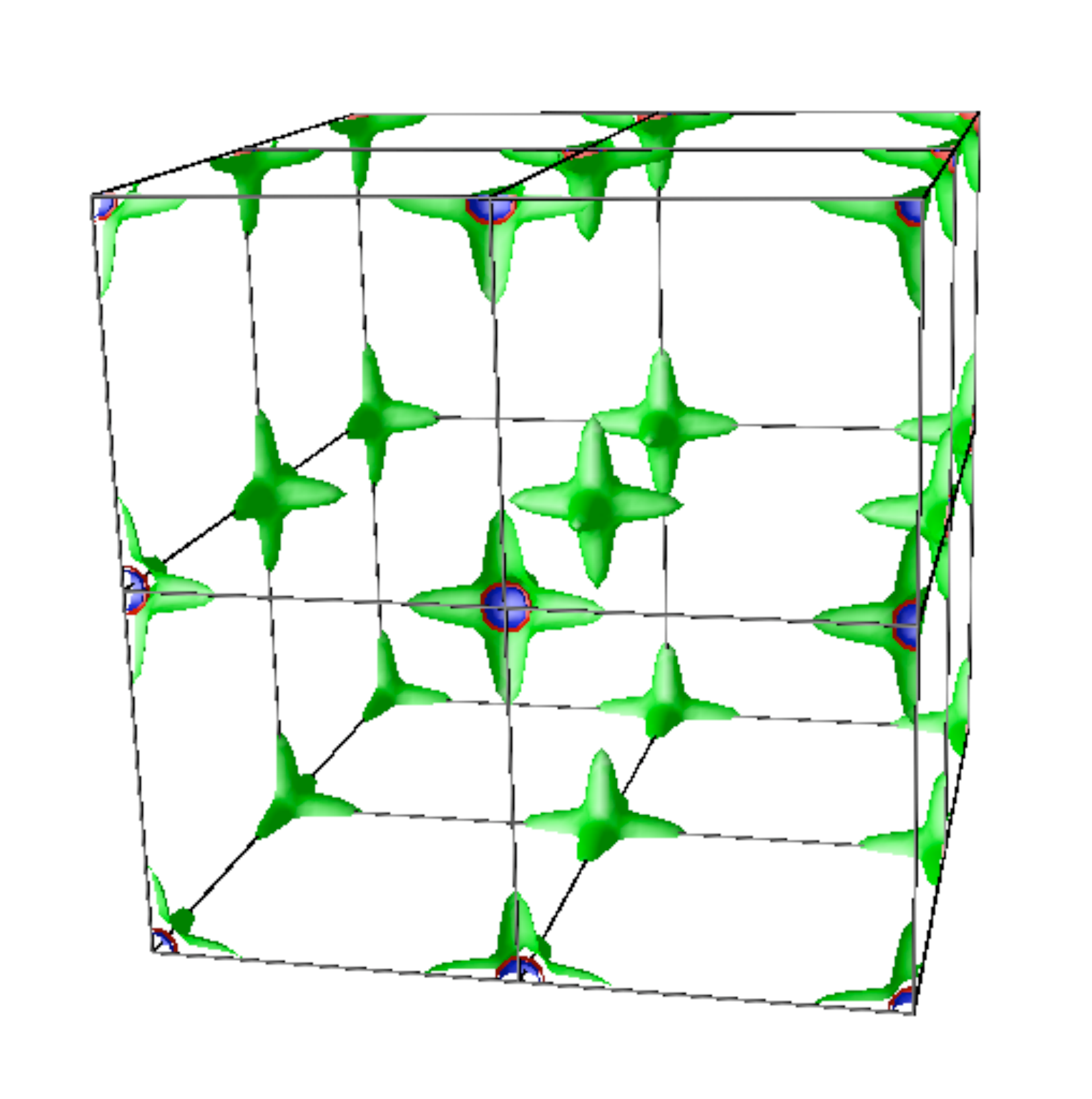}
\caption{Fermi surface of the high-temperature cubic phase at 2\% doping, showing the large anisotropy of the lowest band. At the critical doping $x_c=0.097$ a topological transition takes place where the Fermi surfaces open up along the three axis.}
\label{fig:fs}
\end{center}
\end{figure}
A detailed view of the bandstructure around the zone center is shown in Fig.~\ref{fig:dispersion}. In this limited region of $k-$space the band structure can be effectively described by a tight-binding model within the $t_{2g}$ manifold of the Ti-3d states. The main aspects of the band structure are described by Bloch-waves of $d_{xy}$, $d_{yz}$ or $d_{zx}$ character, each of which has two directions of strong dispersion ($k_x$ and $k_y$ for the $d_{xy}$ orbital etc.) and one slowly dispersing direction orthogonal to these. The result is a set of three degenerate bands. The Fermi surface consists of three interpenetrating ellipsoids centered at the zone center, with the ellipsoids oriented along the x,y and z axis of the reciprocal lattice of the cubic crystal structure. This zone-center degeneracy is however lifted by the spin-orbit interaction. In the cubic phase this results in two degenerate spin-orbit doublets at the lowest energy, and an additional doublet at 29.2 meV higher energy. This splitting equals $1.5\xi$, where $\xi=18.8$ meV is the spin-orbit parameter, somewhat smaller than $\xi=25$ meV used by Mattheiss. In the low temperature tetragonal phase the crystal field $D=2.2$ meV lifts the degeneracy between the two doublets causing a splitting of 4.3 meV. The result is the following set of bands having their minimum energy at the zone center: The lowest "heavy electron" band consists of states carrying angular momentum $m_j = \pm 3/2(1-\delta)$, where $\delta \propto D^2/\xi^2$. While the band disperses upward rather sharply at the zone center, it is deflected downward at $|\vec{k}|\approx 0.1/a$ for momentum along the Ti-O bond. The second band is a "light electron" band, which becomes occupied at the critical carrier concentration $x_{c1}=4.0\cdot 10^{-5}$. Its dispersion is to a good approximation an isotropic parabola, and these bands have the peculiarity that the gyromagnetic factor $g_j=0$ due a compensation of orbital ($g_l=1,m_l=\pm 1$) and spin magnetic moment ($g_s=2,m_s=\pm 1/2$). The third band is also a light electron band which becomes occupied at the critical carrier concentration $x_{c2}=2.6\cdot 10^{-3}$ (n=$4.4\cdot 10^{19}$cm$^{-3}$). An experimental indication for this critical carrier concentration comes form the observation by Binnig {\em et al.}\cite{binnig1980} of an additional superconducting gap of smaller size than the main gap for doping concentrations in excess of $5\cdot 10^{19}$cm$^{-3}$, using tunneling spectroscopy.

\begin{figure}[ht]
\begin{center}
\includegraphics[width=\columnwidth]{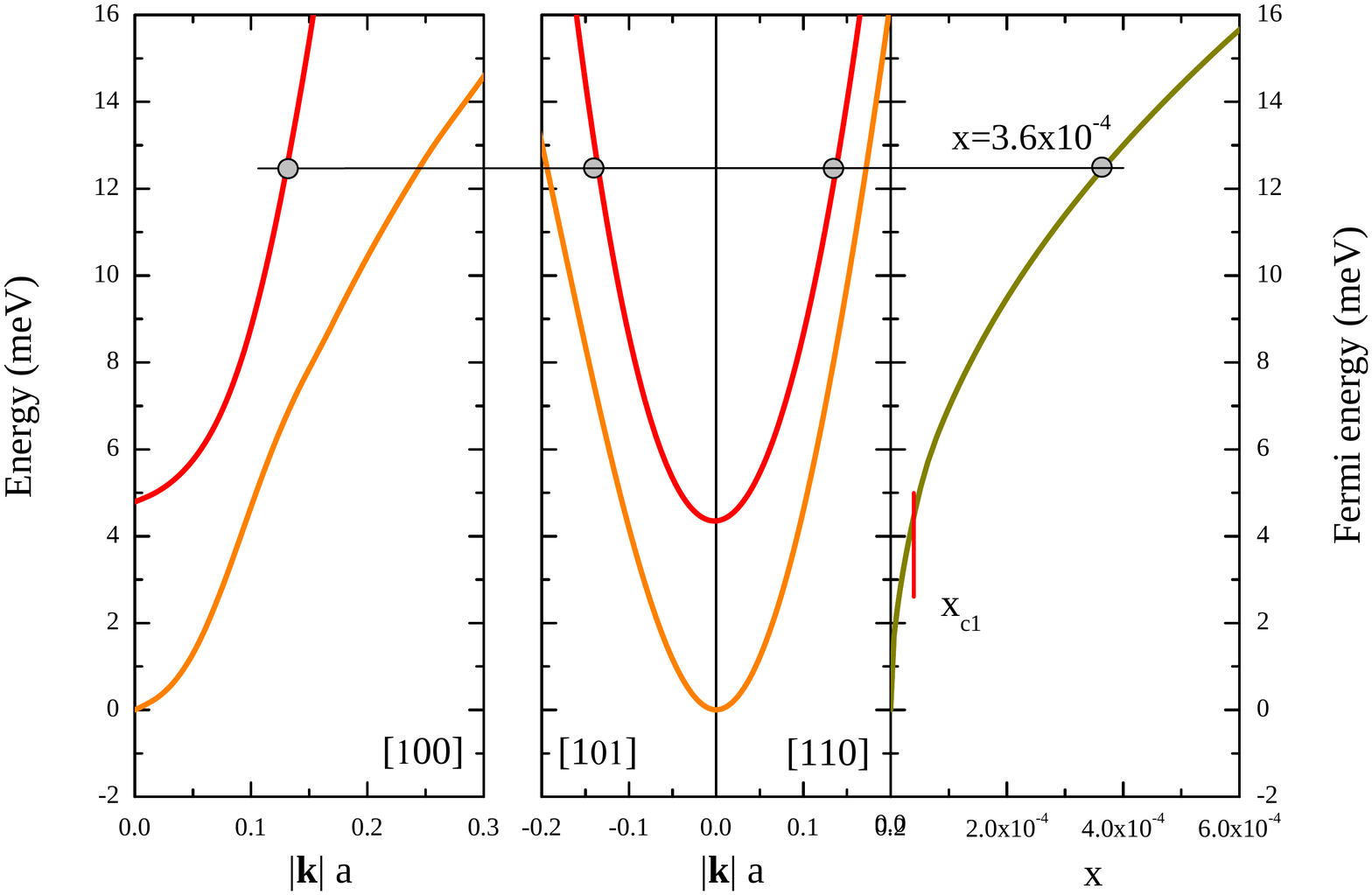}
\caption{Enlarged view of Fig.~\ref{fig:dispersion} indicating the position of the Fermi level for $x=3.6\cdot 10^{-4}$ charge carriers. The corresponding value $ka=0.134$ is in excellent agreement with the hitherto unexplained de Haas-van Alphen frequency reported by Gregory {\em et al.}\cite{gregory1977}. }
\label{fig:dispersion_detail}
\end{center}
\end{figure}

The most signicifant differences between the results presented here and Matheiss' results\cite{mattheiss1972b} are the much smaller crystal field parameter $D=2.2$ meV obtained here as compared to $D=-33$ meV obtained from a tight-binding fit to Matheiss' bands, and the fact that the sign is opposite. The resulting Fermi surface of the lowest band is therefore quite different: In the present calculation it is in fact similar to Fermi-surface of the cubic phase shown in Fig.~\ref{fig:fs} (taking 2\% doping), and has 6 arms extending along $[100]$,$[010]$ and $[001]$. The arms along the z-axis are slightly longer than those along x and y, but on the scale of Fig.~\ref{fig:fs} this is not a perceptible difference. In contrast, Mattheiss's Fermi-surfaces (Fig.~6 of Ref.~\onlinecite{mattheiss1972b}) have 4 arms along $[100]$ and $[010]$ and none along $[001]$. Gregory~\textit{et al.}~\cite{gregory1977} studied samples with electron density $6\cdot 10^{18}$ cm$^{-3}$, corresponding to $x=3.6\cdot 10^{-4}$. Due to the large crystal field splitting, the Fermi level in Mattheiss's calculation is then still below the second band. Yet Gregory~\textit{et al.} observed low-frequency quantum oscillations with frequencies 40 Tesla and 45 Tesla. The weak field-orientation dependence indicated that these are associated with rather isotropic Fermi surfaces, which they associated with the light-electron band. To have this band occupied they postulated that Mattheiss' estimate of the splitting of the two lowest bands introduced by the tetragonal distortion needed to be revised downward. Looking now at our calculation we notice that, since $x=3.6\cdot 10^{-4}>x_{c1}$, the light-electron band is indeed occupied for this doping level. As shown in Fig.~\ref{fig:dispersion_detail}, the diameter of the second Fermi surface is practically independent of direction for this low doping range with a radius $k=0.134/a=3.54\cdot 10^6$ cm$^{-1}$ and extremal area $A=3.95\cdot 10^{13}$ cm$^{-2}$. Using the Onsager relation $F=A \hbar/(2\pi e)$ the corresponding quantum oscillation frequency is 41 Tesla. Since these samples consisted of many domains with different orientation of the tetragonal axis, a doublet due to the anisotropy is expected and observed. It thus appears, that the new {\em ab initio} band structure settles an old conundrum regarding the quantum oscillations of n-type SrTiO$_3$.

\section{Mass renormalization induced by electron-phonon coupling\label{sec:massrenormalization}}
\begin{figure}[ht]
\begin{center}
\includegraphics[width=\columnwidth]{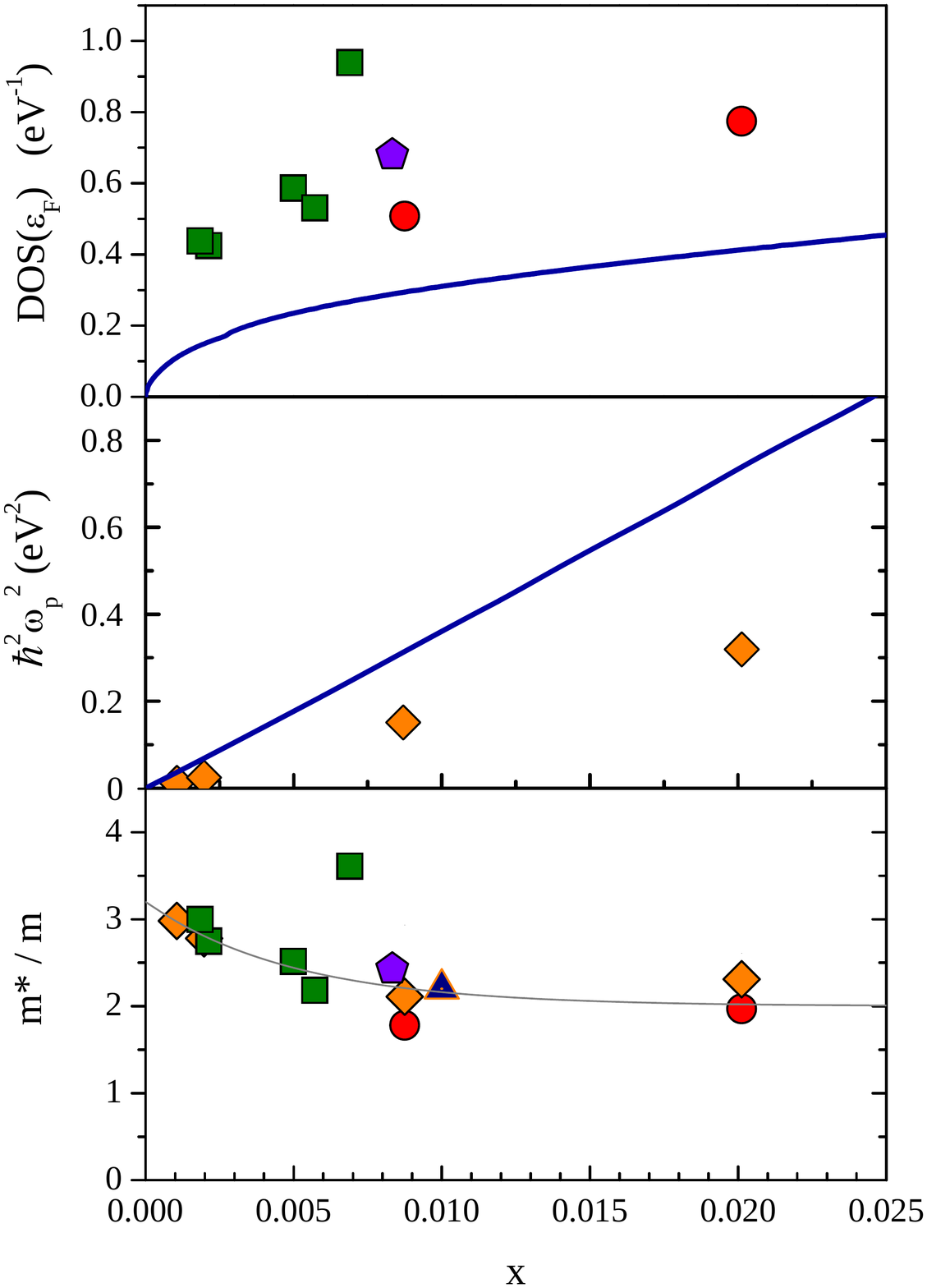}
\caption{Top panel: Doping dependence of the density of states (DOS) at the Fermi energy. The solid curve corresponds to the tight-binding bandstructure fitted to the Wien2k {\em ab initio} bandstructure, with parameters of the first row of table~\ref{table:TB}. Squares\cite{phillips1971}, circles\cite{rolflortz,thesisdook2010}, pentagon\cite{ambler1966}: Density of states obtained from the linear term in the specific heat.
Middle panel: Doping dependence of the Drude spectral weight, $\omega_{p}^2$. The solid curve corresponds to the bandstructure results. Diamonds are the experimental values \cite{vanmechelen2008}.
Bottom panel: (i) Ratio of experimental DOS over bandstructure DOS (experimental and theoretical values taken from top panel, the meaning of the symbols is the same). (ii) Ratio of bandstructure over experimental $\omega_{p}^2$ (values taken from middle panel). (iii) Ratio of bare and experimental (dressed) Fermi-velocity, $v_{F,b}/v_{F,e}$ (triangle, data from Ref.~\onlinecite{meevasana2010}). The grey curve is a smooth interpolation, $m^*/m_b=2.0+1.2\exp{(-x/0.005)}$. }
\label{fig:DOS}
\end{center}
\end{figure}

In Ref.~\onlinecite{vanmechelen2008} we compared the Drude spectral weight to the same quantity calculated using LDA. The expression for the spectral weight along the $x_j$-axis is
\begin{equation}
\omega_{p,b,j}^2=\frac{4\pi e^2}{\hbar^2 V}\sum_{k\nu\sigma} f(\epsilon_{k\nu\sigma}) \frac{\partial^2 \epsilon_k}{\partial k_j^2}
\end{equation}
where the sum is over momentum, band-index and spin and $f(\epsilon)$ is the Fermi-Dirac distribution. The index $b$ in $\omega_{p,b,j}$  refers to the fact that, since the LDA-calculation does not take into-account electron-phonon interaction, it calculates the bare mass. The ratio $\omega_{p,b,j}^2/\omega_{p,e,j}^2$ where $\omega_{p,e,j}^2$ is the experimental Drude spectral weight, then corresponds to the mass renormalization factor $m^*/m_b$. This procedure was followed in Ref.~\onlinecite{vanmechelen2008}. Since the linear term of the specific heat is a direct measure of the density of states at the Fermi energy, $\gamma=\frac{k_B^2\pi^2}{3} N_F$, the ratio $\gamma_{e}/\gamma_{b}$ of the experimental over the LDA value provides a second way to measure the mass enhancement. In Fig.~\ref{fig:DOS} the LDA-calculation of the DOS at $\epsilon_F$ is plotted as a function of doping, together with values obtained from experimental specific heat data. Clearly the DOS as given by experiments is about a factor 2 higher than the LDA-prediction. The corresponding mass enhancement together with the results of the other two methods are summarized in Fig.~\ref{fig:DOS}. The verdict is clear: There is a factor of 2 to 3 mass enhancement with a tendency to become smaller for higher doping. Electron-phonon coupling is the only plausible suspect for the enhancement. Indeed, recent calculations confirm this\cite{devreese2010}: Based on the Fr\"ohlich interaction the essential characteristics of the observed optical conductivity spectra of SrTi$_{1-x}$Nb$_x$O$_3$, in particular intensity, lineshape and energy of a peak at 130 meV, was explained without any adjustment of material parameters. The electron-phonon coupling coupling constant was found to be of intermediate strength.

\begin{figure}[ht]
\begin{center}
\includegraphics[width=\columnwidth]{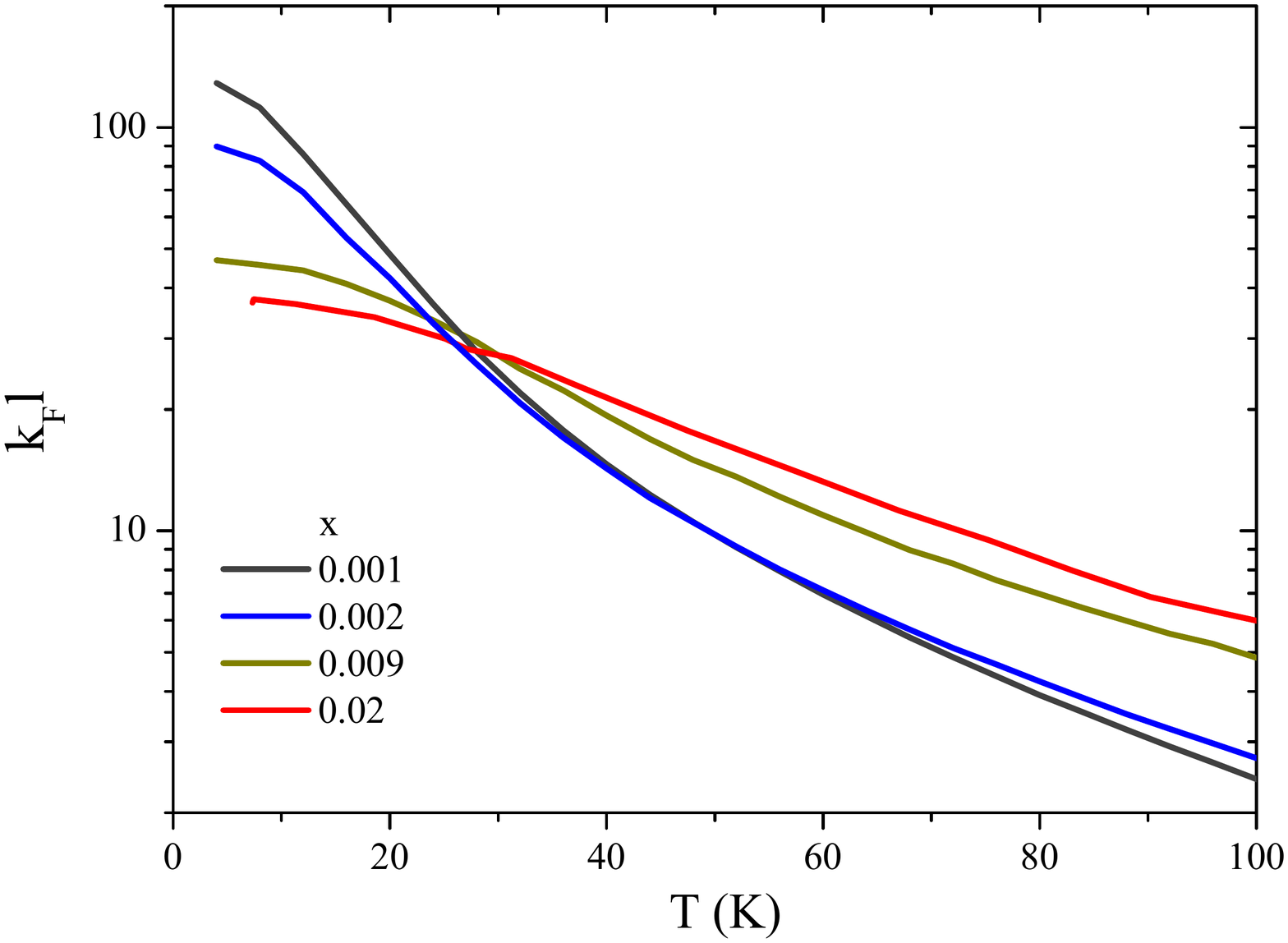}
\caption{The quantity $k_F l$ where $l$ is the mean free path, calculated using $k_F l=2\epsilon^*_F\hbar^{-1}\tau$, and using the experimental $\tau$ of Fig.~\ref{fig:tau} and the calculated Fermi energies corrected by the mass renormalization of Fig.~\ref{fig:DOS}. The high values of $k_F l$ below 100 K imply the itinerant character of the charge carriers.}
\label{fig:ioffe_regel}
\end{center}
\end{figure}

For the correct understanding of the peculiar temperature dependence it is important to find out whether or not the charge carriers are to a good approximation described by Bloch waves. This corresponds to the requirement that the mean-free path at the Fermi-level, $l=v^*_F\tau$ is much bigger than the Fermi-wavelenth, in other words $v^*_F\tau \gg 2\pi/k_F$. Multiplying both sides of the expression with $k_F /2$ we obtain $\epsilon^*_F\hbar^{-1}\tau\gg \pi$, where $\epsilon^*_F/\epsilon_F=v^*_F/v_F=m_b/m^*$. In the previous section we have obtained the doping dependence of $\epsilon_F$. Combining this with the $m^*$ of Fig.~\ref{fig:DOS} and $\hbar/\tau$ of Fig.~\ref{fig:tau} we obtain $k_Fl$ as a function of temperature for different dopings, shown in Fig.~\ref{fig:ioffe_regel}. We see from this graph that at 4 K the electrons are strongly Bloch-like. At low temperatures the largest $k_Fl$ occurs for the lowest carrier concentration. This is expected since in these samples the number of charge carriers is equal to the number of Nb-ions, which act both as donor atoms and scattering potentials. The opposite trend occurs above the isosbectic point (24 K, $k_Fl=33$). While at  4 K we obtain high values of $k_Fl$ in the range from 40 to 150, at 100 K we have $k_Fl$ of the order $2\pi$ implying localization in Fermi-wavelength sized wavepackets. In the low temperature range it is therefore reasonable to extract interaction parameters from the coefficients of the $T^2$ dependence of $1/\tau$. Above approximately 100 K the material enters into a regime of incoherent transport. We therefore restrict the analysis of $1/\tau$ in section~\ref{sec:manybody} to temperatures below 100 K.

\section{T$^2$ relaxation rate and two-body interactions\label{sec:manybody}}
The low carrier density leads to a situation where the kinetic energy of the charge carriers is slaved to the relevant vibrational energy scale. The usual Migdal-Eliashberg expansion in the electron-phonon coupling constant is therefore not applicable. A different approach is required whereby in first instance the electron-phonon coupling is treated for each individual electron, resulting in charge carriers renormalized by electron-phonon coupling, which condense into a Fermi-liquid of "polarons". The polarons interact with each other via the Coulomb interaction and by virtual exchange of phonons. With regards to optical phonons the hierarchy of energy scales is inverted as compared to the situation in common metals, in that $\hbar\omega > \epsilon_F^*$ where $\omega$ is the optical phonon frequency and $\epsilon_F^*$ the Fermi energy of the polarons. While it is clear that the Migdal-Eliashberg expansion can not be used, the solution of the many-body problem in this limit is a complicated problem which we will not attempt to solve here. Instead we turn the problem around and anticipate that the correct solution should share certain properties in common with the problem of interacting composite fermions such as $^3$He. The essential properties should then be those of fermions interacting through some effective interaction mediated by the optical phonons, which on the scale of $\epsilon_F^*$ can be considered effectively non-retarded. An immediate consequence is then a $T^2$ contribution to the inelastic relaxation rate, resulting from the phonon-mediated fermion-fermion interaction\cite{lawrence1973,lawrence1976,macdonald1980}

The situation in SrTi$_{1-x}$Nb$_x$O$_3$ is more complicated than in $^3$He in that the Fermi surface is crossed by 3 bands for $x>2.6 \cdot 10^{-3}$ and by 2 bands for $4.0 \cdot 10^{-5}<x<2.6  \cdot 10^{-3}$. However, we assume that the transport and superconducting behavior are dominated by the most highly occupied band, and use the single-band expressions in Appendix~\ref{appendix:B} to derive the effective coupling constants.

If indeed the temperature dependence of the relaxation rate is a manifestation of fermion-fermion scattering, it should be possible to obtain from it a parameter characterizing the interaction strength. The formalism has been elaborated in the context of the observation of van Kempen {\em et al.} of a T$^2$ contribution in the resistivity of potassium\cite{vanKempen1976}. In particular, the expression for the relaxation rate is (see appendix~\ref{appendix:B})
\begin{eqnarray}
\frac{\hbar}{\tau} &=& a_2 T^2 \nonumber \\
 a_2 &=& \lambda_{\tau}^2u\frac{\pi^3 k_B^2}{\epsilon^*_F}
\label{eq:a2}
\end{eqnarray}
The parameter $u\le 1$ describes the fraction of the momentum changes which is transferred to the ionic lattice.

The dimensionless parameter $\lambda_{\tau}$ represents the interaction effective in polaron-polaron scattering. Since we have determined $a_2$ and $\epsilon^*_F$ in the previous sections, we are ready to calculate $\lambda_{\tau}^2u$ using this expression. The result is shown in the lower panel of Fig.~\ref{fig:g_Tc}.
The trend of $\lambda_{\tau}u^{1/2}$ going to zero in the zero-doping limit may be a consequence of either $\lambda_{\tau}$ or $u$ diminishing in the low doping limit, or a combination of these two. As pointed out in Appendix~\ref{appendix:B}, several factors make $u\neq 0$: (i) Baber scattering involving heavy and light electrons (ii) Umklapp scattering, and/or (iii) disorder scattering by donor atoms. Since for $u\rightarrow 0$ (i) the mass-anisotropy (section \ref{sec:bands}), (ii) the probability of intra-pocket Umklapp, and (iii) the impurity scattering $1/\tau_0$ (see Table Ref.~\ref{table:drude}) all vanish, we may expect $u \rightarrow 0 $ in this limit. Since we will see in the following section that the polaron-polaron interaction as calculated from T$_c$ is almost independent of doping, we tentatively attribute the observed doping dependence of $\lambda_{\tau}^2u$ to a suppression of $u$ for low doping.

\section{Superconductivity\label{sec:superconductivity}}
Superconductivity is observed in n-type SrTiO$_3$, with a dome shaped T$_c$ between 0 and 0.02 electrons per SrTiO$_3$ formula unit \cite{koonce1967}, with maximum values of about 0.7 K when doped with Nb\cite{binnig1980}. This doping dependence and the T$_c$ itself are relatively robust features of the doped 3-dimensional bulk materials as well as the 2-dimensional SrTiO$_3$/LaAlO$_3$ interfaces \cite{reyren2007,caviglia2008}. Superconductivity in doped bulk SrTiO$_3$ has been anticipated by M. L. Cohen on the basis of an attractive electron-electron interaction arising from the exchange of intravalley and intervalley phonons\cite{cohen1964} and motivated by early bandcalculations\cite{kahn1964} indicating a many-valley bandstructure in SrTiO$_3$. The intervalley mechanism has been further elaborated in the context of SrTiO$_3$ in a number of papers\cite{eagles1967,eagles1969,koonce1969}. However, over the years evidence has been accumulating that all bands are at the center of the Brillouin zone. Several alternative mechanism {\em not} involving a multi-valley bandstructure have been proposed, to mention a few: (i) J. Appel\cite{appel1969} noticed that the Brillouin-zone folding associated with the tetragonal distortion creates two bands of zone-folded optical phonons with a quasi-accoustic dispersion at the zone center. One of these bands has a finite matrix element for intravalley scattering and can therefor in principle mediate superconducting pairing. (ii) Z. Zinamon, while maintaining Appel's idea of soft phonon exchange, argued that the relevant charge carriers are small polarons, and proposed a theoretical model relevant for this limit\cite{zinamon1970}, (iii) T. Jarlborg has demonstrated by electronic structure calculations that the electron-phonon coupling is enhanced for long-wavelength phonon mode despite the low density of states, which is consistent with the appearance of superconductivity at low doping\cite{jarlborg2000}. We see, that the mechanism for pairing in this material is far from clear. We therefor adopt here a phenomenological approach whereby we deduct the coupling constant characterizing pairing interaction from the experimental T$_c$'s and compare it to the coupling constant obtained from the transport data. While this approach does not solve the question as to the exact nature of the phonon-mediated interaction, it does allow to establish whether superconductivity and transport properties can be treated in a unified approach of an interacting Fermi-liquid.

One of the consequences of the Fermi energy being smaller than the relevant phonon energy $\omega_c$, is that the energy cutoff of the pairing interaction is given by $\epsilon^*_F$ on the occupied side of the Fermi level, while it is given by $\omega_c$ on the unoccupied side. This introduces a dependence of T$_c$ on $\epsilon_F^*$ which may in part be responsible for the decrease of T$_c$ for $x\rightarrow 0$.
\begin{figure}[ht]
\begin{center}
\includegraphics[width=\columnwidth]{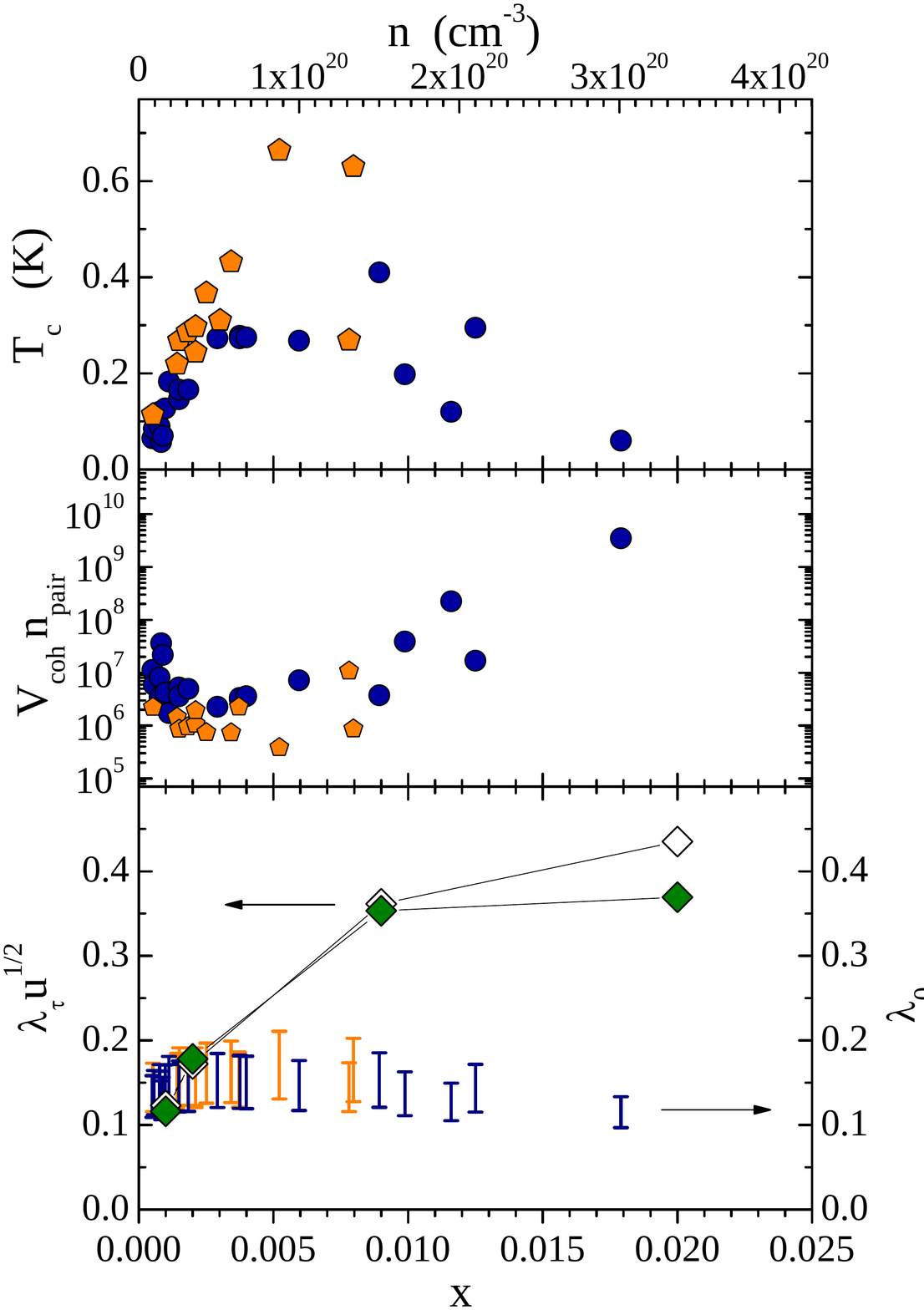}
\caption{Top panel: T$_c$ of a large number of samples. Pentagons: data by Binnig {\em et al.}\cite{binnig1980}. Circles: data by Koonce {\em et al.}\cite{koonce1967}. Middle panel: Coherence volume times the density of Cooper pairs. The high value indicates that each pair is overlapping with a huge number of other pairs. Lower panel: Doping dependence of the coupling constants $\lambda_{\tau}$ obtained from the $T^2$ term of the resistivity and $\lambda_0$ from T$_c$. Circles: $\lambda_{\tau}$ using $a_2$ listed in the second column of table~\ref{table:fit} and Eq.~\ref{eq:a2}. Diamonds: {\em idem} using $a^{\prime}_2$. Bars: coupling constants $\lambda$ from $T_c$ at the corresponding carrier concentration of SrTi$_{1-x}$Nb$_x$O$_3$ using the T$_c$'s of the top panel and Eq.~\ref{gapequation2}, $\epsilon^*_F=\epsilon_Fm_b/m^*$ using $\epsilon_F$ from Fig.~\ref{fig:dispersion} and $m^*/m_b$ from Fig.~\ref{fig:DOS}. Upper (lower) limits of the bars indicate the value for $g$ obtained with $\omega_c=$ 5 meV (80 meV).}
\label{fig:g_Tc}
\end{center}
\end{figure}
A more detailed discussion of the consequences in the limit of weak coupling ($T_c\ll T_F$, as is the case in all samples that we discuss here) is provided in Appendix~\ref{appendix:C}.
Before we set out to discuss this, it is important to establish whether the superconductivity is closer to the BCS limit, or to the limit of Bose-Einstein condensation of bipolarons. The latter has been proposed for low carrier concentrations ($n<10^{18}$cm$^{-3}$) in Zr doped SrTiO$_3$\cite{eagles1969} and for the high Tc cuprates \cite{alexandrov2011}. One way to investigate this question is, by estimating with how many other pairs each Cooper-pair overlaps. In the BEC-limit there is essentially no overlap, whereas in a BCS superconductor it is given by the volume occupied by a pair divided by the available volume. The former is just $(4\pi/3)\xi_0^3$ where $\xi_0=\hbar v_F/\pi\Delta_0$, and the latter is $2/n$ where $n$ is the electron density. Using standard relations between density and Fermi energy we obtain
\begin{equation}
\frac{V(occupied)}{V(available)}=\frac{4}{9\pi^4}\left(\frac{\epsilon^*_F}{\Delta_0}\right)^3
\end{equation}
Binnig {\em et al.}~\cite{binnig1980} have observed gap values close to $\Delta_0/k_BT_c=1.76$ in their tunneling spectra for exactly the data in Fig.~\ref{fig:g_Tc}, hence we can use this substitution for $\Delta_0$. The result is shown in the middle panel of Fig.~\ref{fig:g_Tc} in a broad doping range using data collected by Koonce {\em et al.}~\cite{koonce1967} and Binnig {\em et al.}~\cite{binnig1980}. Ipso facto each Cooper-pair overlaps with 10$^5$ to 10$^{10}$ others, which places these superconductors clearly outside the realm of Bose-Einstein condensation for the range of carrier concentrations considered here.
Substituting in Eq.~\ref{gapequation2} of Appendix~\ref{appendix:C} the values of T$_c$ and the value of $\epsilon^*_F$ discussed in section~\ref{sec:massrenormalization} we calculate the corresponding coupling constant $\lambda_0$ for the pairing interaction. Since we lack certainty about the nature and frequency of the phonons causing the pairing interaction, we have substituted two extremal values for the vibrational cutoff-energy in the gap equation: $\omega_c$=5 meV and 80 meV. The resulting uncertainty of $\lambda_0$ is not very large; for all dopings we find $0.1<\lambda_0<0.2$ with negligible doping dependence. The results are shown in Fig.~\ref{fig:g_Tc} together with $\lambda_{\tau}u^{1/2}$. The different doping dependence of $\lambda_0$ and $\lambda_{\tau}u^{1/2}$ has a simple explanation in that we expect the parameter $u$ to vanish for $x\rightarrow 0$.

\section{Landau parameters\label{sec:fermiliquid}}
\begin{figure}[ht]
\begin{center}
\includegraphics[width=\columnwidth]{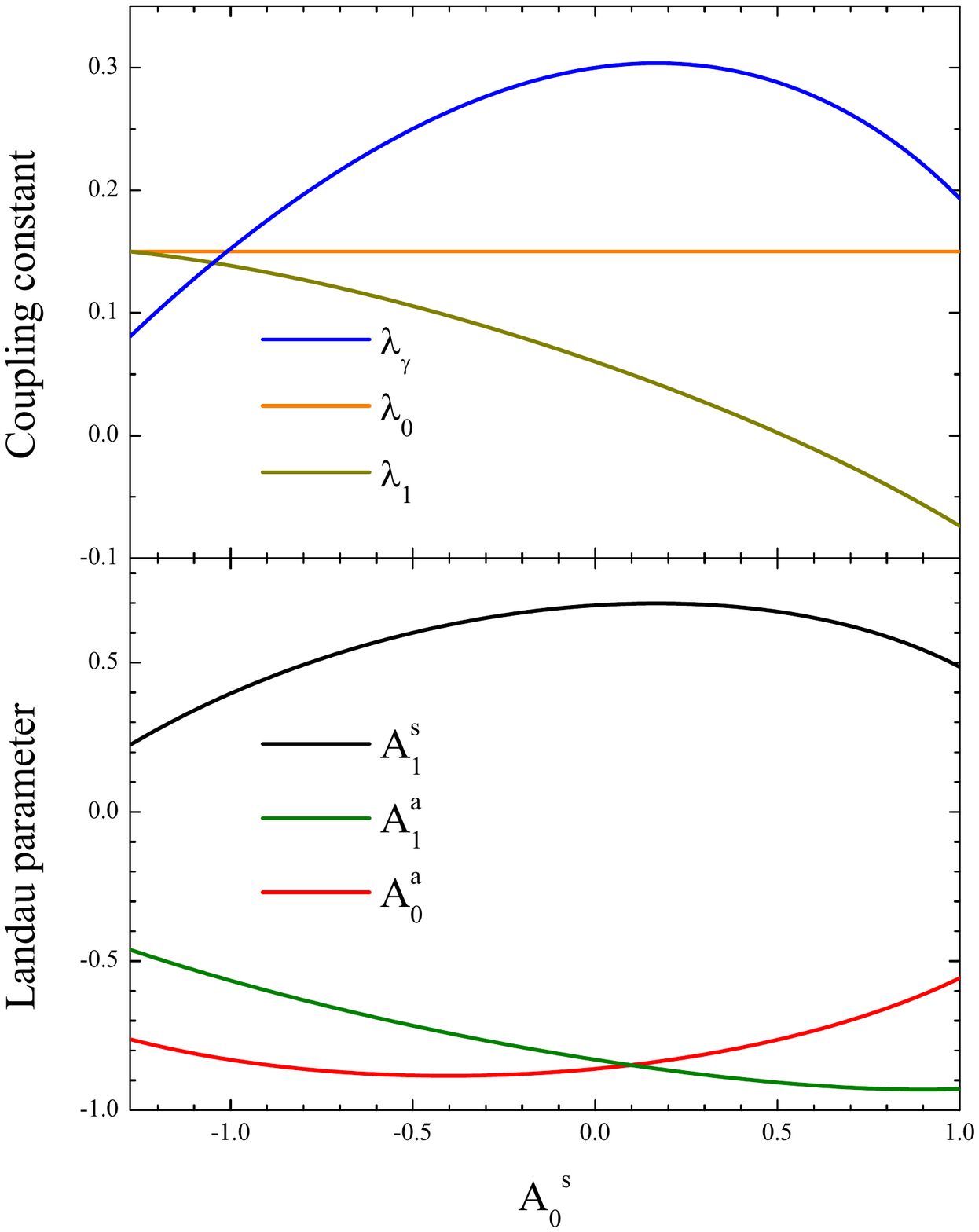}
\caption{Solutions for $A_1^s$, $A_0^a$, $A_1^a$, the triplet superconducting coupling constant and the mass-enhancement as a function of $A_0^s$ when taking the experimentally determined $\lambda_{\tau}=0.38$, $\lambda_0=0.15$ and the BPR sum rule as constraints assuming singlet pairing for the ground state. }
\label{fig:Landau}
\end{center}
\end{figure}

In principle we want to determine the full set of relevant Landau parameters, either in the form $A_l^j$ or as $F_l^j$. However, even while we have assumed that the only relevant angular momentum values are $l=0,1$, there are still 4 parameters while until now we have determined two quantities which depend on them, namely $\lambda_0$ and $\lambda_{\tau}u^{1/2}$. The first question concerns the symmetry of the pairing itself: The expression relating the superconducting coupling constant to the Landau parameters are different for singlet and triplet pairing, so one has to make a choice as to whether one assumes triplet or singlet superconductivity. Triplet pairing can be excluded because the only available mechanism in the present case is electron-phonon coupling. The second question concerns the value of $u$. In the previous section we attributed the suppression of $\lambda_{\tau}u^{1/2}$ as $x\rightarrow 0$ to the suppression of momentum transfer to the ionic lattice. Vice versa, in Fig. \ref{fig:g_Tc} we see that $\lambda_{\tau}u^{1/2}$ has saturated for $x > 0.2$. Accuracy by which the Umklapp fraction can be calculated is probably within a factor of 2, even in the alkali metals which are relatively simple due to the nearly free electron character\cite{bass1990,wiser1992}. We make the simplest possible assumption that $u\approx 1$, implying that for the higher doping levels $\lambda_{\tau}\approx 0.4$. We will base the analysis of the Landau parameters on this value. With 4 parameters to determine and 2 experimental constraints we need two additional pieces of information. One of them is supplied by the sum rule for the Landau parameters derived by Brinkman, Platzman and Rice\cite{brinkman1968} (BPR sum rule) for charged fermions, which for the $sp$-model implies
\begin{equation}
 A_1^s + A_0^a + A_1^a= -1
\label{eq:sum rule}
\end{equation}
We are still one constraint short. One might hope to find such a constraint in, for example, the mass-enhancement measured with specific heat. The problem is however that one needs to compare the values of the electronic specific heat with and without polaron-polaron interactions. Since the mass of a polaron is already enhanced compared to the bare band mass by a factor of 2 approximately, to extract the contribution of polaron-polaron interactions, especially if it is much smaller than 1 as it turns out to be the case here, is difficult and at the present state of affairs not feasible. We therefore calculated $A_0^a$, $A_1^s$, and $A_1^a$ as a function of $A_0^s$ while fixing the constraints imposed by $\lambda_0=0.15$ through Eq. \ref{eq:gst}, by $\lambda_{\tau}=0.4$ through Eq. \ref{eq:gtau} and the sum rule Eq.\ref{eq:sum rule}. The parameter $A_0^s$ is varied in the range of positive mass enhancement ($m^*/m-1=F_1^s/3>0$), and positive compressibility ($\kappa=\kappa_b(1-A_0^s)m^*/m>0$, where $\kappa_b$ is the bare value). Since we assume that the pairing symmetry is of the singlet variety, we only consider the solutions for $\lambda_0>\lambda_1$. The result shown in Fig. \ref{fig:Landau} allows to determine all parameters once the value of $A_0^s$ has been set. The difference $A_1^a-A_0^a$ represents an exchange interaction which tends to align spins parallel for positive values. Its value increases for $A_0^s\rightarrow -1.27$. Correspondingly, for $-1.27<A_0^s<-1.15$ an alternative set of solutions is obtained corresponding to triplet pairing (not displayed in the figure) which, as already pointed out above, we reject on theoretical grounds.

Even with this broad range of possibilities for $A_0^s$ allowed by the experimental constraints, the windows for $A_0^a$, $A_1^s$, and $A_1^a$ are limited. In table \ref{table:Landau} we compare all parameters discussed here to the case of $^3$He at ambient pressure. We see, that the values for SrTiO$_3$ are of the same order, but smaller than in liquid $^3$He. The fact that we obtain "reasonable", i.e. not excessively large or small numbers, of the Landau parameters, gives further support to the notion that that n-type SrTiO$_3$ is a Landau Fermi liquid, and superconductivity and the T$^2$ resistivity in this compound have a common origin.

\begin{table}[ht]
\begin{tabular}{c|c|c}
 \hline
  Parameter&SrTiO$_3$& $^3$He,1 atm.\\
    &singlet& triplet\\
 \hline
  A$_0^s$	& $\{-1.27 ; 1.0\}$	  &0.91	 \\
  A$_1^s$	& $0.45\pm0.25$	      &2.0	   \\	
  A$_0^a$	& $-0.67\pm0.22$	  &-2.03 \\
  A$_1^a$	& $-0.62\pm0.28$	  &-0.55	\\
 \hline					
\end{tabular}
\caption{Ranges of the Landau parameters of SrTiO$_3$ allowed by the constraints imposed by the experimental data combined with the BPR sum rule (2d column, this work) and values of the same of $^3$He (3d column, see Ref.~\onlinecite{mahan2000}, original data in Refs.~\onlinecite{wheatley1975,feenberg1969}).}
\label{table:Landau}
\end{table}
\section{Conclusions}
We have performed a detailed analysis of the T$^2$ behavior of the resistivity of n-type SrTiO$_3$. Novel bandstructure data are presented, and it is shown that the new band structure solves an old conundrum of the de Haas-van Alphen frequencies dating from 1977. The mass, density of states, and plasma frequency of the quasiparticles are found to be renormalized by approximately a factor 2 due to electron-phonon interaction. The quasiparticles turn out to be in the anti-adiabatic limit with respect to electron-phonon interaction with a quasi-instanteneous interaction mediated through phonons. Analysis of the $T^2$ resistivity and $T_c$ provides values of the Landau parameters of n-type SrTiO$_3$ which are comparable in size to those of liquid $^3$He.
\section{Acknowledgements}
This work is supported by the SNSF through Grant No. 200020-130052 and the National Center of Competence in Research (NCCR) "Materials with Novel Electronic Properties-MaNEP". We thank R. Lortz for assistance with the specific heat experiments. DvdM acknowledges stimulating discussions with J. Devreese, S. Gariglio, A. Georges, T. Giamarchi, D. Jaccard, M. M\"uller, and J.-M. Triscone. We are grateful to J. Levallois and A. B. Kuzmenko for their comments on the manuscript.

\appendix
\section{Bandstructure\label{appendix:A}}
Calculations have been performed in the high-temperature perovskite (HTP) structure ($a=$3.905 \AA ) as well as in the low-temperature tetragonal (LTT) structure (group no. 140, I4/mcm, $a=$5.529 \AA , $c=7.824$ \AA , O$_{x}=0.244$); optimizing the O position in the calculations yields O$_{x}=0.223,$ indicating, not surprisingly, that even zero-point fluctuations substantially reduce the average distortion. Nb doping was simulated in the virtual crystal approximation, changing the nuclear charge of Ti from 22 to $22+x$, or that of Sr from 38 to $38+x$ (the results did not change, proving that this is a good approximation in the considered range of dopings).

Most calculations were performed with $RK_{\max }=7$ for the wave-function expansion, and $RG_{\max }=14$ for the charge density expansion; calculations with $RK_{\max }=8$ and $RG_{\max }=16$ did not show a discernable difference. Fermi-surface integrals were evaluated using the k-point meshes up to $28\times 28\times 28.$ The plasma frequencies were evaluated as the Fermi-surface averages of the squared Fermi velocities. The Fermi velocities were calculated using the WIEN2k optics package; accuracy of the Fermi velocities was tested (in the cubic case) by numerical differentiation of the energy eigenvalues.

For many calculations of physical properties it is useful to have an integration in momentum space which is rapid. We therefore used a parametrization of the {\em ab initio} band structure of the $t_{2g}$ manifold in the region around the zone center, based on the following tight-binding model
\begin{eqnarray}
\epsilon_{\vec{k},j}=4t_{\pi}\sum_{i\neq j}\sin^2\left(\frac{k_ia}{2}\right)+4t_{\delta}\sin^2\left(\frac{k_ja}{2}\right)
\nonumber\\
H_{k}=
\left(\begin{matrix}
\epsilon_{\vec{k},1}&0&0\\
0&\epsilon_{\vec{k},2}&0\\
0&0&\epsilon_{\vec{k},3}
\end{matrix}\right)
%
%
+\frac{1}{2}
\left(\begin{matrix}
2D & \xi &  \xi \\
\xi &  2D&  \xi\\
\xi  &  \xi &  -4D
\end{matrix}\right)
\end{eqnarray}
with the corresponding parameters given in table~\ref{table:TB}.
\begin{table}[ht]
\begin{tabular}{ccccc}
  \hline
  Source &$t_{\delta}$&$t_{\pi}$&$\xi$&$D$\\
         &meV&meV&meV&meV\\
  \hline
  Wien2k code  &35 & 615 & 18.8 & 2.2 \\
  Mattheiss  &35 & 500 & 28 & -33 \\
  \hline
\end{tabular}
\caption{Tight-binding parameters describing the dispersion of the t$_{2g}$ bands near the $\Gamma$-point of SrTiO$_3$. First row: Parameters fitted to the Wien2k code bandstructure presented here. Second row: Parameters fitted to Matheiss' calculations\cite{mattheiss1972b} for the tetragonal phase with a tilt angle of 2.1$^o$.}
\label{table:TB}
\end{table}

\section{Landau parameters\label{appendix:B}}
In the Landau-Fermi liquid theory of interacting fermions, the bare interaction is expressed in terms of the dimensionless Landau parameters $F_l^j$ where the index $l$ indicates the angular momentum and $j$ the parity of the scattering process. The mass renormalization depends only on $F_1^s$
\begin{equation}
\lambda_{\gamma}=\frac{m^{*}}{m} -1= \frac{F_1^s}{3}
\end{equation}
The specific heat is renormalized by the same factor as the effective mass. The interaction between dressed quasiparticles
\begin{equation}
A_l^j=\frac{F_l^j}{1+F_l^j/(2l+1)}
\end{equation}
is important for scattering between quasiparticles and constitutes the pairing interaction for superconductivity.
Following Dy and Pethick\cite{dy1969} we limit the scattering processus to the $l=0$ and $l=1$ values in the spherical expansion of the scattering amplitudes (the so-called $s-p$ approximation)
\begin{eqnarray}
A_s(\theta,\phi) &=& \frac{1}{N_F}\left[(A_0^s-3A_0^a)+(A_1^s-3A_1^a)\cos\theta\right] \\
A_t(\theta,\phi) &=& \frac{1}{N_F}\left[(A_0^s+A_0^a)+(A_1^s+A_1^a)\cos\theta\right]\cos\phi \nonumber
\end{eqnarray}
where $N_F=m^*k_F/(\pi^2\hbar^2)$ is the density of states at the Fermi level and the angles $\theta$ and $\phi$ represent the kinematics of the scattering events.
The relevant quantity in the theory of inelastic scattering is the transition probability $W(\theta,\phi)$\cite{nozieres1999,mahan2000}
\begin{equation}
W(\theta,\phi)=\frac{\pi}{4\hbar}
\left[A_s(\theta,\phi)+A_t(\theta,\phi)\right]^2+\frac{\pi}{2\hbar}A_t(\theta,\phi)^2
\end{equation}
Due to collisions between quasiparticles, the relaxation rate of the dressed quasiparticles has a $T^2$ temperature dependence, which is the most characteristic property of a Landau Fermi liquid. We will follow here the approach of Lawrence and Wilkins\cite{lawrence1973,lawrence1976,macdonald1980}. These authors define a surface-averaged relaxation rate for an electron at the Fermi surface\cite{tau0}, and derive the relation between $\tau$ and $W(\theta,\phi)$
\begin{equation}
\frac{1}{\tau}= \frac{(m^*)^3(k_B T)^2u}{12\pi^2\hbar^6}
\left\langle\frac{W(\theta,\phi)}{\cos{(\theta/2)}}\right\rangle
\label{eq:tau_W}
\end{equation}
where the dimensionless coefficient $u<1$ represents the efficiency of momentum transfer to the ionic lattice of the relaxation process. In a translationally invariant system of interacting electrons $u=0$, because the current operator commutes with the Hamiltonian of such a system. However, the fact that a solid does not possess full translation symmetry has important consequences. Already in 1937 Baber demonstrated a mechanism for finite resistivity in a two-band model in which $s$ electrons are scattered from heavier $d$ holes by a screened Coulomb interaction\cite{baber1937}. The Baber mechanism works more generally for a system of light and heavy electrons with the heavy particles acting as momentum sinks, and, as pointed out Giamarchi and Shastry, similar results are expected for typical, noncircular bands\cite{giamarchi1992}. This last point is relevant to the case of SrTiO$_3$ in view of the strong $k$-dependence of the mass across the Fermi-surface (see Fig. \ref{fig:fs}). In single band Umklapp processes allow momentum transfer to the crystal coordinate system\cite{lawrence1973}. Likewise, the potential landscape caused by impurities (for example the donor and/or acceptor atoms in doped semiconductors) provides a channel by which momentum gets transferred to the ionic lattice in electron-electron collisions. Refs.~\onlinecite{lawrence1973,lawrence1976,macdonald1980} concentrated on alkali-metals, and therefor did not take into account Baber scattering. They used the symbol $\Delta$ for the Umklapp fraction. To avoid confusion with the superconducting gap, we indicate here the fraction of momentum in electron-electron collisions transferred to the ionic lattice due to Umklapp and other mechanisms with the character $u$.
Integration of the angular integrals in Eq.~\ref{eq:tau_W} is straightforward though tedious, with the result
\begin{equation}
\left\langle
 \frac{W(\theta,\phi)}{\cos{(\theta/2)}}
\right\rangle
=12\lambda_{\tau}^2\frac{\pi^5\hbar^5}{(m^*)^3\epsilon^*_F}
\label{eq:W}
\end{equation}
where
\begin{eqnarray}
12\lambda_{\tau}^2=
\frac{7}{24}(A_1^s)^2
+\frac{49}{40}(A_1^a)^2
-\frac{7}{20}A_1^sA_1^a&\nonumber\\
+\frac{5}{8}(A_0^s)^2
+\frac{21}{8}(A_0^a)^2
-\frac{3}{4}A_0^sA_0^a&\nonumber\\
-\frac{5}{12}A_0^sA_1^s
-\frac{7}{4}A_0^aA_1^a
+\frac{1}{4}A_0^sA_1^a
+\frac{1}{4}A_0^aA_1^s&
\label{eq:gtau}
\end{eqnarray}
If $A_1^s$ is the only non-zero parameter we obtain $\lambda_{\gamma}/(1+\lambda_{\gamma})=4/3 \lambda_0=4\sqrt{2/7}\lambda_{\tau}$. Finally, by substituting Eq.~\ref{eq:W} in Eq.~\ref{eq:tau_W} we obtain
\begin{equation}
\frac{\hbar}{\tau}= \frac{\pi}{\epsilon^*_F}\lambda_{\tau}^2u(\pi k_B T)^2
\label{eq:tau_T}
\end{equation}
This is the central expression enabling extraction of $\lambda_{\tau}^2u$ from the experimental values of the amplitude of the $T^2$ term in the resistivity.
\section{T$_c$ equation\label{appendix:C}}
The gap equation for an isotropic gap is
\begin{equation}
1=\int_{-\infty}^{\infty}d\epsilon
\frac{N(\epsilon)V(\epsilon)}{2\sqrt{(\epsilon-\mu)^2+\Delta^2}}
\tanh\left(\frac{(\epsilon-\mu)^2+\Delta^2}{2k_BT}\right)
\end{equation}
The chemical potential $\mu(T_c)$ is to be determined at the critical temperature by adjusting it such as to fix the number of electrons
\begin{equation}
\int_{0}^{\infty} N(\epsilon) \frac{1}{1+e^{\beta_c(\epsilon-\mu)} } d\epsilon =\int_{0}^{\epsilon_F^*}N(\epsilon)d\epsilon \nonumber
\end{equation}
As it turns out to be the case for the data considered in the present paper, $k_BT_c << \epsilon_F^*$; consequently the output of the self-consistent solution is $\mu(T_c) \approx \epsilon_F^*$. The critical temperature is obtained by solving the gap equation for $\Delta=0$. In the present case the bottom of the band constitutes the lower limit of the integral over the density of states. We define it as the zero of energy, so that
\begin{equation}
1=\int_{0}^{\infty}d\epsilon N(\epsilon)V(\epsilon)\frac{\tanh(\beta_c(\epsilon-\mu)/2)}{2(\epsilon-\mu)}d\epsilon
\label{gapequation1}
\end{equation}
where  $k_BT_c=1/\beta_c$. The usual approximation for the retarded interaction consists of substituting $N(\epsilon)V(\epsilon)=\lambda$ for $|\epsilon-\mu| < \omega_c$ where $\omega_c$ is cutoff energy of the pairing interaction, and taking $\lambda=0$ for $|\epsilon-\mu|> \omega_c$. The expression for $T_c$ is then
\begin{eqnarray}
k_BT_{c,j}=1.13\omega_c \exp\left(\frac{-1}{\lambda_{j}}\right)
\end{eqnarray}
where $\omega_c$ is the cutoff energy of the pairing interaction, and the coupling constants for the $l=0$ (singlet) and $l=1$ (triplet) pairing channels are\cite{patton1975}
\begin{eqnarray}
\lambda_{0}&=&\frac{1}{4}\left(A_1^s-A_0^s\right)+\frac{3}{4}\left(A_0^a-A_1^a\right)\nonumber\\
\lambda_{1}&=&\frac{1}{12}\left(A_1^s-A_0^s\right)-\frac{1}{12}\left(A_0^a-A_1^a\right)
\label{eq:gst}
\end{eqnarray}
In the case of n-type SrTiO$_3$ an interesting asymmetry is introduced by the condition that the energy scale of the phonons mediating the interaction is in the anti-adiabatic limit: we have $\mu< \omega_c$. Since on the occupied side $N(\epsilon) = 0$ for $-\omega_c < \epsilon  - \mu < - \mu$, the region of finite $N(\epsilon)V(\epsilon)$ is limited to $-\mu < \epsilon - \mu < \omega_c$. Another aspect to take into account is that $N(\epsilon)=c\sqrt{\epsilon}$. We therefore define the dimensionless coupling constant $\lambda$ at $\mu$ through the relation $N(\epsilon)V(\epsilon)=\lambda\sqrt{\epsilon/\mu}$ and we make a transformation of variables $x=\beta_c(\epsilon - \mu)$. The equation for T$_c$ then becomes
\begin{equation}
\frac{1}{\lambda}=\frac{1}{\sqrt{\beta_c\mu}}\int_{-\beta_c\mu}^{\beta_c\omega_c}
\sqrt{x+\beta_c\mu}
\frac{\tanh(x/2)}{2x}dx
\label{gapequation2}
\end{equation}
which has the following solution in the weak coupling limit ($\lambda<1$)
\begin{equation}
T_c=0.612\mu
\exp{\left(\sqrt{\frac{\omega_c }{\mu}}\right)}\exp{\left(\frac{-1}{\lambda}\right)}
\label{eq:Tclambda}
\end{equation}
The values for $\lambda_0$ shown in Fig.~\ref{fig:g_Tc} were obtained by solving Eq.~\ref{gapequation2} numerically, and agree within 3\% accuracy with the weak coupling expression Eq.~\ref{eq:Tclambda} for $\omega_c$=80 meV.
\end{document}